\documentclass[aps,prd,
showpacs,superscriptaddress,groupedaddress]{revtex4}  
 
\usepackage{graphicx}  
\usepackage{dcolumn}   
\usepackage{bm}        
\usepackage{amssymb}   

\usepackage{amsmath,amsfonts,
latexsym,
} 
\usepackage{slashed} 
\usepackage{relsize} 
\usepackage{prodint}

\begin{document}



\title{Precanonical structure of the Schr\"odinger wave 
 functional 
 in curved space-time}

\author{Igor V.  Kanatchikov}
\affiliation{School of Physics and Astronomy, University of St Andrews, St Andrews KY16 9SS, UK} 
\email {ik25@st-andrews.ac.uk} 
\affiliation{National Quantum Information Center in Gda\'nsk (KCIK), 81-831 Sopot, Poland  }




\date{\today}


\begin{abstract}
A relationship between the functional Schr\"odinger representation and the precanonical quantization of a scalar field theory is extended to arbitrary curved space-times. The canonical functional derivative Schr\"odinger equation is derived from the manifestly covariant precanonical Schr\"odinger equation and the Schr\"odinger wave functional is expressed as the trace of the product integral of Clifford-algebra-valued precanonical wave functions restricted to a certain field configuration when the ultraviolet parameter $\varkappa$ introduced in precanonical quantization is infinite. Thus the standard QFT in functional Schr\"odinger representation emerges 
from the precanonical formulation of quantum fields as a singular limiting case. 

\end{abstract}



\noindent

\maketitle


\newcommand{\beq}{\begin{equation}}
\newcommand{\eeq}{\end{equation}}
\newcommand{\beqa}{\begin{eqnarray}}
\newcommand{\eeqa}{\end{eqnarray}}
\newcommand{\nn}{\nonumber}
\newcommand{\bew}{\begin{widetext}}
\newcommand{\eew}{\end{widetext}}

\newcommand{\half}{\frac{1}{2}}

\newcommand{\xt}{\tilde{X}}

\newcommand{\uind}[2]{^{#1_1 \, ... \, #1_{#2}} }
\newcommand{\lind}[2]{_{#1_1 \, ... \, #1_{#2}} }
\newcommand{\com}[2]{[#1,#2]_{-}} 
\newcommand{\acom}[2]{[#1,#2]_{+}} 
\newcommand{\compm}[2]{[#1,#2]_{\pm}}

\newcommand{\lie}[1]{\pounds_{#1}}
\newcommand{\co}{\circ}
\newcommand{\sgn}[1]{(-1)^{#1}}
\newcommand{\lbr}[2]{ [ \hspace*{-1.5pt} [ #1 , #2 ] \hspace*{-1.5pt} ] }
\newcommand{\lbrpm}[2]{ [ \hspace*{-1.5pt} [ #1 , #2 ] \hspace*{-1.5pt}
 ]_{\pm} }
\newcommand{\lbrp}[2]{ [ \hspace*{-1.5pt} [ #1 , #2 ] \hspace*{-1.5pt} ]_+ }
\newcommand{\lbrm}[2]{ [ \hspace*{-1.5pt} [ #1 , #2 ] \hspace*{-1.5pt} ]_- }

\newcommand{\pbr}[2]{ \{ \hspace*{-2.2pt} [ #1 , #2\hspace*{1.4 pt} ] 
\hspace*{-2.3pt} \} }
\newcommand{\nbr}[2]{ [ \hspace*{-1.5pt} [ #1 , #2 \hspace*{0.pt} ] 
\hspace*{-1.3pt} ] }

\newcommand{\we}{\wedge}
\newcommand{\nbrpq}[2]{\nbr{\xxi{#1}{1}}{\xxi{#2}{2}}}
\newcommand{\lieni}[2]{$\pounds$${}_{\stackrel{#1}{X}_{#2}}$  }

\newcommand{\rbox}[2]{\raisebox{#1}{#2}}
\newcommand{\xx}[1]{\raisebox{1pt}{$\stackrel{#1}{X}$}}
\newcommand{\xxi}[2]{\raisebox{1pt}{$\stackrel{#1}{X}$$_{#2}$}}
\newcommand{\ff}[1]{\raisebox{1pt}{$\stackrel{#1}{F}$}}
\newcommand{\dd}[1]{\raisebox{1pt}{$\stackrel{#1}{D}$}}
\newcommand{\der}{\partial}
\newcommand{\oo}{$\Omega$}
\newcommand{\Om}{\Omega}
\newcommand{\om}{\omega}
\newcommand{\eps}{\epsilon}
\newcommand{\si}{\sigma}
\newcommand{\Lm}{\bigwedge^*}

\newcommand{\inn}{\hspace*{2pt}\raisebox{-1pt}{\rule{6pt}{.3pt}\hspace*
{0pt}\rule{.3pt}{8pt}\hspace*{3pt}}}
\newcommand{\sro}{Schr\"{o}dinger\ }
\newcommand{\vol}{\omega}
               \newcommand{\dvol}[1]{\der_{#1}\inn \vol}

\newcommand{\bd}{\mbox{\bf d}}
\newcommand{\bder}{\mbox{\bm $\der$}}
\newcommand{\bI}{\mbox{\bm $I$}}

\newcommand{\gammat}{\; \widetilde{\gamma}{}}
\newcommand{\be}{\beta} 
\newcommand{\ga}{\gamma} 
\newcommand{\de}{\delta} 
\newcommand{\Ga}{\Gamma} 
\newcommand{\gmu}{\gamma^\mu}
\newcommand{\gnu}{\gamma^\nu}
 \newcommand{\ka}{\varkappa} 
 \newcommand{\la}{\lambda}
\newcommand{\hka}{\hbar \kappa}
\newcommand{\al}{\alpha}
\newcommand{\lapl}{\bigtriangleup}
\newcommand{\psib}{\overline{\psi}}
\newcommand{\Psib}{\overline{\Psi}}
\newcommand{\Phib}{\overline{\Phi}}
\newcommand{\derts}{\stackrel{\leftrightarrow}{\der}}
\newcommand{\what}[1]{\widehat{#1}}

\newcommand{\gammab}{\bar{\gamma}}

\newcommand{\bx}{{\bf x}}
\newcommand{\bk}{{\bf k}}
\newcommand{\bq}{{\bf q}}

\newcommand{\omk}{\omega_{\bf k}} 
\newcommand{\lpl}{\ell}
\newcommand{\zb}{\overline{z}} 

\newcommand{\deltab}{\bar \delta}

\newcommand{\dv}{\mbox{\sf d}}

\newcommand{\deltt}{\bm{\delta}}   

\newcommand{\BPsi}{\mathbf{\Psi}} 
\newcommand{\BPhi}{\mathbf{\Phi}}
\newcommand{\BH}{{\bf H}} 
\newcommand{\BS}{{\bf S}} 
\newcommand{\BN}{{\bf N}} 
\newcommand{\BXi}{{\bf \Xi}}

\newcommand{\rd}{\mathrm{d}}
\newcommand{\ri}{\mathrm{i}}
\newcommand{\Tr}{\mathrm{Tr}} 

\newcommand{\boldx}{{\bx}} 
\newcommand{\fnc}{{\rm function}}
\newcommand{\equn}{{\rm equation}}
\newcommand{\ota}{{\rm obtain}}
\newcommand{\dwh}{{\rm DW Hamiltonian}}
\newcommand{\fd}{{\rm field}}
\newcommand{\con}{{\rm connection}}
\newcommand{\bewn}{{\rm between}}
 \newcommand{\vsh}{{\rm vanish}}
 \newcommand{\lmt}{{\rm limit}}
 \newcommand{\tmm}{{\rm term}}
 \newcommand{\cse}{{\rm case}}
 \newcommand{\Sch}{{\rm Schr\"{o}dinger}}
 \newcommand{\wnn}{{\rm when}}
 \newcommand{\ltr}{{\rm latter}}
 \newcommand{\rsd}{{\rm restricted}}
  \newcommand{\drv}{{\rm derivat}}
 \newcommand{\clf}{{\rm Clifford}}
\newcommand{\fst}{{\rm first}}
\newcommand{\rln}{{\rm relation}}
\newcommand{\dfrt}{{\rm different}}
\newcommand{\rsp}{{\rm respect}}
\newcommand{\by}{{\rm by}}
\newcommand{\spt}{{\rm space-time}}
\newcommand{\crsp}{{\rm correspond}}
\newcommand{\trt}{{\rm treat}}
\newcommand{\hlt}{{\rm Hamilton}} 
\newcommand{\and}{{\rm and}}
\newcommand{\qnt}{{\rm quant}} 
\newcommand{\myy}{{\rm we}}
\newcommand{\imls}{{\rm implies}}
\newcommand{\dnt}{{\rm denote}}
\newcommand{\gma}{\gamma}
\newcommand{\Si}{\Sigma}


\section{Introduction}  

Since the theoretical discovery of the Hawking radiation of black holes, quantum field theory in curved space time \cite{qft-curved} often has been considered as an opportunity to study the interplay between gravitation, space-time and quantum theory with a view to gaining  insights and intuitions into the quantum geometry of space-time and the quantum theory of gravity.  The consideration of quantum fields on curved backgrounds also allows us to understand what concepts and mathematical structures are important in quantum field theory 
beyond the simplifying framework of the Poincar\'e-invariant Minkowski spacetime. 

Recently we have put forward an approach to quantization of fields called precanonical quantization  
\cite{berlin,bl,lz,gequ} 
which is based on mathematical structures of the De Donder-Weyl (DW) Hamiltonian theory known in the 
calculus of variations \cite{dewe}.  This Hamiltonian-like formulation 
does not require a space-time decomposition and treats all space-time variables on equal footing. 
In this sense it is an intermediate description of classical fields between the Lagrangian and the canonical Hamiltonian one 
(hence the name "precanonical") 
 which allows us to avoid the necessity of treating fields as infinite-dimensional 
Hamiltonian systems 
and the technical difficulties it brings in at least on the level of formulating the quantum theory.  
In DW Hamiltonian theory the Poisson brackets are defined on the dynamical variables represented by differential forms 
rather than functionals, and they lead  to the structure of Poisson-Gerstenhaber algebra \cite{gequ,pg1,pg2} 
(see also \cite{jo1,jo2,loday,rmr,dkp00,dirbr} for further generalizations). 
 The latter generalizes the  Poisson algebra in the canonical Hamiltonian formalism which underlines canonical quantization.

The DW Hamiltonian formulation is related to a  generalization of the Hamilton-Jacobi (HJ) theory which is formulated in terms 
of partial derivative rather than functional derivative equations \cite{dewe}. 
Precanonical quantization clarifies the question as to which formulation of quantum theory of fields reproduces the 
DWHJ  equation in the classical limit \cite{bl,lz}. We found that quantization of a small Heisenberg-like subalgebra of the aforementioned Poisson-Gerstenhaber structure leads to a hypercomplex generalization of the formalism of quantum theory where both operators and wave functions are Clifford-algebra-valued. The precanonical analogue of the  Schr\"odinger equation is formulated using the Dirac operator on the space-time which appears as a multidimensional generalization of the time derivative in the left side of the standard Schr\"odinger equation \cite{berlin,bl,lz,gequ}. 
 
One of the features of the precanonical formulation of quantized  fields is that it allows us to reproduce the classical field equations in DW Hamiltonian form as the equations of expectation values of operators defined precanonical quantization and evolving according to the precanonical Schr\"odinger equation \cite{my-ehr,mg-ehr}. By treating the space-time variables on equal footing it leads to a formulation of quantum theory of fields on a finite-dimensional space of field and space-time variables thus  
providing  a promising framework for the quantum gauge theory \cite{romp2018,ym-mass}  
and the theory of quantum gravity \cite{pqgr,qg}. 

In order to be able to realize the potential of precanonical quantization we have to demonstrate how it could be related to more 
familiar and already working concepts of standard QFT.  In this paper we continue our previous work where a relationship between 
precanonical quantization and the functional Schr\"odinger picture in QFT has been established \cite{my-pla,my-ym1,atmp1,atmp2,romp2018,static18} and extend the results obtained there to general curved space-times. 


\section{Quantum scalar field on a curved space-time: canonical and precanonical description}

{ Let us recall that the conventional {\bf canonical quantization} of scalar field theory in curved space-time 
can be formulated in the functional Schroedinger representation of QFT  \cite{hatf,jacw}.
It leads \cite{frs,pi,shor1,shor2,cor1,cor2,kief}) to the description of the corresponding quantum field in terms 
of the Schr\"odinger wave functional $\BPsi([\phi(\bx)],t)$  satisfying the 
Schr\"odinger equation 
\bew
\beq \label{fs}
\ri\hbar \der_t \BPsi = 
 \int \! d\bx\, \sqrt{-g}
 \left (  \frac{\hbar^2}{2}\ \frac{g_{00}}{g}\frac{\delta^2}{\delta \phi(\bx)^2} 
- \frac{1}{2} g{}^{ij} \der_i\phi(\bx)\der_j\phi(\bx) + V(\phi)
 \right) \BPsi , 
\eeq 
\eew 
where 
the right hand side is the canonical Hamiltonian operator formulated in terms of functional 
derivative operators, $x^\mu = (t,\bx)= (t, x^i)$ 
are space-time coordinates,  $g_{\mu\nu}$ is the space-time metric tensor whose components depend on $x^\mu$, 
$g = \det \Vert g_{\mu\nu}\Vert$. In (\ref{fs}) 
one uses the space-time coordinates adapted to 
the space-like foliation such as  the induced metric on the space-like leaves of the foliation is $g_{ij}$,   
the lapse $N = \sqrt{g_{00}}$ and the shift functions  $N_i = g_{0i}=0$.

The {\bf precanonical quantization} of a scalar  field $\phi(x)$ on a curved space-time  background 
given by  the metric tensor $g_{\mu\nu} (x)$   (cf.  \cite{my-ehr,mg-ehr}) 
 gives rise to the description  in terms of a  wave function $\Psi(\phi, x^\mu)$ 
 on the finite-dimensional bundle with the coordinates $(\phi,x^\mu)$ 
which takes values in the complexified space-time Clifford algebra, i.e. 
\beq 
\Psi = \psi + \psi_\mu\gamma^\mu + \frac{1}{2!} \psi_{\mu_1\mu_2}\gamma^{\mu_1\mu_2}+...
+ \frac{1}{n!}\psi_{\mu_1...\mu_n}\gamma^{\mu_1...\mu_n} , 
\nn
\eeq
and satisfies 
the partial derivative precanonical Schr\"odinger equation (pSE)
\bew
\beq  \label{crv-ns}
\ri\hbar \gamma^\mu (x) 
 \nabla_\mu \Psi = 
 \left(- \frac{1}{2} \hbar^2\varkappa \frac{\der^2}{\der \phi^2 } 
+ \frac1\ka V(\phi)  \right)\Psi =: \frac1\ka\what{H}\Psi \,, 
\eeq
 \eew
where 
$\gamma^\mu (x)$ are the  curved space-time Dirac matrices such that 
\beq \label{dima}
\gamma^\mu (x)\gamma^\nu (x) + \gamma^\nu (x) \gamma^\mu (x) = 2g^{\mu\nu}(x), 
\eeq 
$\gamma^{\mu_1...\mu_p}$ are the antisymmetrized products of $p$ Dirac matrices, 
\beq
\nabla_\mu := \der_\mu + \frac14\omega_\mu(x)
\eeq 
is the covariant derivative  with the spin-connection matrices 
$\omega_\mu (x) = \omega_\mu{}_{AB}(x)\underline{\gamma}{}^{AB}$ 
 (see e.g. \cite{pol}) acting on Clifford-algebra-valued wave functions 
 by the commutator product \cite{static18}, 
 and 
$\underline{\gamma}{}_{A}$ are the 
constant Dirac matrices which factorize the 
Minkowski metric $\eta_{AB}$ of the tangent space\footnote{We chose the signature 
 $+---...$. Note that in this paper we depart from our notation in \cite{mg-ehr} 
where the flat Dirac matrices were denoted $\gammab{}^A$. This notation can be confusing when 
the Dirac conjugate matrix has to be denoted as $\overline{\gammab}{}^A$.  As for the rest, 
throughout this paper we mostly follow the notation and conventions used in 
\cite{my-ehr,mg-ehr,atmp1,atmp2}.}: 
\beq 
\underline{\gamma}{}_{A} \underline{\gamma}{}_{B} 
+ \underline{\gamma}{}_{B} \underline{\gamma}{}_{A} = 2 \eta_{AB}. 
\eeq

The operator $\what{H}$ 
in (\ref{crv-ns}) is the 
De Donder-Weyl (DW) Hamiltonian operator which is constructed according to the procedure of 
precanonical quantization \cite{bl,lz,my-ehr,mg-ehr}. In the expression of $\what{H}$ there appears 
an ultraviolet parameter $\ka$ of the dimension of the inverse spatial volume.  
This parameter typically appears in the representations of precanonical 
quantum operators  \cite{berlin,bl,lz,my-ehr}. 
For the scalar fields on curved background the DW Hamiltonian operator $\what{H}$ 
coincides with its couterpart in flat space-time (cf. \cite{berlin,bl,lz,my-ehr}).  
Correspondingly,  
the curved space-time  manifests itself  only through the curved space-time Dirac matrices (\ref{dima}) 
and the spin-connection term on the left-hand side of (\ref{crv-ns}).


 Obviously, the description of quantum fields obtained from precanonical quantization is very different from  a familiar description of quantum fields derived from the canonical quantization. In particular, while in the description using the functional Schr\"odinger picture  the role of space variables $\bx$ 
  is different from the role  the time variable $t$, 
the precanonical description is entirely space-time symmetric, manifestly covariant and independent on the assumption of global hyperbolicity of space-time. One can also wonder how the description in terms of precanonical wave function on a finite dimensional space and the corresponding PDE can possibly match the description in terms of functionals on an infinite-dimensional space of field configurations at a fixed time and the corresponding functional derivative Schroedinger equation,  or how the multiparticle states and multi-point correlation functions of standard QFT could be related to the natural objects within the precanonical description such as the Green function of the precanonical Schr\"odinger equation (\ref{crv-ns}).  
 
However, one can reduce the percepted gap between those two description by noticing that already on the classical level the solutions 
 of field equations can be equally well treated using both the language of PDE on finite dimensional spaces (in the the Lagrangian and 
 DW Hamiltonian and DWHJ descriptions) and the language of functional derivative equations (in the canonical Hamiltonian 
 and Hamilton-Jacobi description). Moreover, one can derive the canonical Hamiltonian and HJ equations from the DW Hamiltonian and 
 DWHJ equations, respectively (see e.g. \cite{my-pla,atmp2}). 
  In the next section we will show how those relationships between 
 canonical and precanonical are extended to the quantum level in curved space-times.

\section{Relating the precanonical wave function and the Schr\"odinger wave functional 
} 

 Our preceding work has established a relationship  between the functional Schr\"odinger 
representation and precanonical quantization of scalar and Yang-Mills in flat space-time
\cite{atmp1,atmp2,romp2018} 
The familiar QFT in functional Schr\"odinger representation 
was derived  from the precanonical quantization as the limiting 
case when the combination $\underline{\gamma}{}_0\ka$ is replaced by  $\delta(\mathbf{0})$, 
a regularized value of Dirac delta function $\delta(\bx-\bx')$ at 
coinciding spatial points, which can be understood as the cutoff of the momentum space volume introduced by a regularization. 
Here we intend to extend this relationship to curved space-time using the example of a quantum scalar  field.

The Schr\"odinger wave functional $\BPsi([\phi(\bx)], t)$ is interpreted as the probability amplitude of finding a field configuration $\phi(\bx)$ at some moment of time $t$. The precanonical wave function $\Phi(\phi,x)$ is the probability amplitude of observing the field value $\phi$ at the space-time point $x$. Then the time-dependent  complex functional probability amplitude $\BPsi([\phi(\bx)], t)$ 
can be expected to be a composition of space-time dependent Clifford-valued probability amplitudes given by the precanonical wave function  $\Psi(\phi,x)$. It means that the  Schr\"odinger wave functional $\BPsi ([\phi(\bx)],t)$ 
is a  functional of precanonical wave functions $\Psi (\phi, x)$
restricted to a specific field configuration  which is represented by a section $\Sigma$ in the total space of the bundle with the 
coordinates $(\phi,x)$,  which is defined by the equation  $\Sigma: \phi=\phi(\bx)$ at time $t$. 
Thus by denoting the  restriction of precanonical wave function $\Psi(\phi,x)$  to $\Sigma$ as 
\beq 
\Psi_\Sigma (\bx,t) := \Psi (\phi=\phi(\bx), \bx, t) \nn 
\eeq 
we assume that 
\beq \label{psib}
\BPsi([\phi(\bx)],t) = \BPsi ([\Psi_\Sigma (\bx,t), \phi (\bx)]) ,   
\eeq 
so that the time dependence of the wave functional $\BPsi$ is totally controlled by the 
time dependence of precanonical wave function restricted to $\Sigma$. Then the chain rule differentiation yields the the 
time derivative of $\BPsi$ 
\beq \label{dtps}
 \ri\der_t \BPsi = 
 {\Tr} \int\! d\bx\, 
 \left \{ 
 \frac{\delta \BPsi }{\delta\Psi^T_\Sigma(\bx, t)} 
\, \ri\der_t \Psi_\Sigma (\bx, t)  
\right \} , 
 \eeq
where $\Psi^T$ denotes the transpose of the matrix $\Psi$. In the following we will avoid unnecessarily cumbersome notation  by denoting 
$\Psi_\Sigma (\bx, t)$  also as $\Psi_\Sigma (\bx)$ or even $\Psi_\Sigma$. 

\subsection{Restriction of precanonical Schr\"odinger equation to $\Sigma$}

The time derivative 
of $\Psi_\Sigma$ is determined by the restriction of pSE 
(\ref{crv-ns}) rewritten in space+time split form 
to  $\Sigma$:  
\begin{widetext}
\beq \label{eq6n}
\ri\der_t \Psi_\Sigma = -\ri\gamma_0\gamma^i \left ( \frac{d}{dx^i} 
- \der_{i} \phi (\bx) \frac{\der}{\der \phi } \right ) \Psi_\Sigma 
- \frac{\ri}{4} \gamma_0 \gamma^{i} [\omega_i{}_\Sigma, \Psi_\Sigma ]  
- \frac{\ri}{4}  [\omega_0{}_\Sigma, \Psi_\Sigma ]
+ \frac{\gamma_0}{\ka} \what{H}_\Sigma \Psi_\Sigma, 
\eeq 
\end{widetext}
where  $\frac{d}{dx^i} $ is the total derivative along $\Sigma$,  
\beq \label{ttl}
\frac{d}{dx^i}:= \der_i + \der_i \phi(\bx)  \frac{\der}{\der \phi} 
+ \der_i \phi_{,k}(\bx) \frac{\der}{\der \phi_{,k} } +... \,.  
\eeq
In (\ref{ttl}) $\phi_{,k}$ denote the fiber coordinates of the 
first-jet bundle of the bundle of field varibles $\phi$ over space-time 
(cf. \cite{snd}) 
and $\what{H}_\Sigma$ is the restriction  of the 
DW Hamiltonian operator $\what{H}$ to $\Sigma$. Since $\what{H}$ contains no space-time derivatives, $\what{H}_\Sigma = \what{H}$ 
and 
\beq \label{dwhop} 
\frac1\ka\what{H} = 
-  \frac{\ka}{2}  \frac{\der^2}{\der \phi^2 } 
+ \frac1\ka V(\phi) . 
\eeq


\subsubsection{Total covariant derivative}

Let us introduce the notion of the total covariant derivative acting on Clifford-algebra-valued 
tensors,  particularly on those restricted to $\Si$. The derivative will be called ``total" in the sense that 
(i) when acting on a Clifford-valued tensor function $T^{\mu_1\mu_2...}_{\nu_1\nu_2...}$  it includes both the spin-connection 
 matrix $\omega_\mu{}^{}$ and the Christoffel symbols $\Gamma^\alpha_{\beta\gamma}$ 
 (c.f. \cite{bertl}) and (ii) when a tensor quantity with the components depending both on 
 $x$ and $\phi$ is restricted to $\Si$, its derivative with respect to $x$-s is understood 
 in the sense of the total derivative (\ref{ttl}): 
\begin{align} \label{cov}
\nabla^{\mathrm{tot}}_\alpha T^{\mu_1\mu_2...}_{\nu_1\nu_2...} := &
\frac{d}{d x^\alpha} T^{\mu_1\mu_2...}_{\nu_1\nu_2...} 
+ \frac14 [\omega_\alpha, T^{\mu_1\mu_2...}_{\nu_1\nu_2...}] 
\nn \\ &+\Gamma^{\mu_1}_{\alpha \beta} T^{\beta \mu_2...}_{\nu_1\nu_2...} 
+ \Gamma^{\mu_2}_{\alpha \beta} T^{\mu_1\beta ...}_{\nu_1\nu_2...}  +.... 
\nn \\&- \Gamma^{\beta}_{\alpha \nu_1} T^{\mu_1 \mu_2 ...}_{ \beta\nu_2...}  
 - \Gamma^{\beta}_{\alpha \nu_2} T^{\mu_1 \mu_2 ...}_{ \nu_1 \beta...}
- ... 
\end{align}
 The commutator 
 in the second term guarantees that the total covariant derivative 
 of the Clifford product of two Clifford-valued tensor quantities fulfills the Leibniz rule. 
 The Christoffel symbols appear in the covariant derivative of non-scalar Clifford quantities, 
 e.g. in the condition of  
 metric compatibility 
\beq  \label{mcomp}
\nabla^{\mathrm{tot}}_\alpha \gamma^{\mu} 
= 0 , 
\eeq 
where only the first partial derivative term in (\ref{ttl}) is non-vanishing  when acting on $x$-dependent $\gamma$-matrices.  

Now, in terms of the total covariant derivative 
$\nabla^{tot}$ acting on $\Psi_\Si$.  eq. (\ref{eq6n})  
takes the form 
\bew
\beq \label{27}
\ri\partial_t \Psi_\Si = -\ri\gamma_0\gamma^i \nabla^{{\mathrm{tot}}}_i \Psi_\Si  
 - \frac\ri4 [\omega_{0\Si}, \Psi_\Si ] 
 + \ri\gma_0\gamma^i\partial_{i} \phi (\bx) \partial_\phi\Psi_\Si 
 +  \frac1\varkappa \gamma_0\widehat{H}_\Si \Psi_\Si .
\eeq
\eew

 \subsection{Time evolution of the Schr\"odinger wave functional from pSE} 
 
From (\ref{dtps}), (\ref{eq6n}) and (\ref{dwhop}) the equation of the time evolution of the wave functional (\ref{psib})
constructed from precanonical wave functions takes the form  
 \bew
\begin{align} \label{dt}
\begin{split}
i\der_t \BPsi =
    \Tr \int \!\rd\bx\ 
 \bigg\{
 \frac{\delta \BPsi }{\delta\Psi^T_\Sigma(\bx, t)} 
\Big[ &
\underbrace{-\ri\gamma_0\gamma^i \frac{d}{dx^i} \Psi_\Sigma  (\bx)}_{I} 
 + \underbrace{\ri \gamma_0\gamma^i\der_i \phi(\bx){\der_\phi} \Psi_\Sigma  (\bx)
 \vphantom{\frac{C}{D}}}_{II}
   \\
 \underbrace{\strut - \frac{\ri}{4} \gamma_0 \gamma^{i} [\omega_i{}_\Sigma, \Psi_\Sigma (\bx)]
 \vphantom{\frac{A}{B}} }_{IIIa}  &
 \underbrace{\strut - \frac{\ri}{4}  [\omega_0{}_\Sigma, \Psi_\Sigma ]}_{IIIb} 
\underbrace{- \frac{\varkappa}{2} \gamma_0 \der_{\phi\phi}\Psi_\Sigma  (\bx)}_{IV} 
+ 
  \underbrace{\frac{1}{\varkappa}\gamma_0 V(\phi(\bx)) \Psi_\Sigma  (\bx)}_{V}
\Big ] \bigg\} . 
\end{split}
\end{align} 
\eew 
  In order to derive from  this equation the functional derivative 
Schr\"odinger equation (\ref{fs}) we need to try to express the terms in the right hand side of (\ref{dt}) in terms of the 
functional derivatives of $\BPsi$ in the form (\ref{psib}) with respect to $\phi(\bx)$.  Those are calculated below.

\subsection{Functional derivatives of $\mathbf{\Psi}$}
By using the chain rule for the functional differentiation
and introducing the notations  
 \beq \label{ffi}
\BPhi^{}({\bx})
:= \frac{\delta \BPsi}{\delta\Psi^T_\Sigma (\bx)}  
\eeq 
and  
\beq 
\der_\phi \Psi_\Sigma(\bx) := ({\der}\Psi  /{\der \phi}) |_\Sigma (\bx) , \quad 
\der_{\phi\phi} \Psi_\Sigma (\bx) := ({\der^2}\Psi /{\der \phi^2})  |_\Sigma (\bx),
\eeq
we obtain 
\bew
\begin{align} 
\label{dbps}
 \frac{\delta \BPsi }{\delta \phi(\bx)}
&=
\Tr\left \{
 \BPhi^{}({\bx})
\der_\phi \Psi_\Sigma (\bx)
\right \}
+  \frac{\deltab \BPsi }{\deltab \phi(\bx)^{{}^{}}}, \\[1ex]
\begin{split}
\label{dbps2} 
\frac{\delta^2 \BPsi }{\delta \phi(\bx)^2}
&=
\Tr\left \{
 \delta(\mathbf{0}) \BPhi^{}({\bx}) \der_{\phi\phi} \Psi_\Sigma (\bx) 
+ 2 \frac{ \deltab \BPhi(\bx)}{\deltab \phi(\bx)}   ~\der_\phi \Psi_\Sigma (\bx) \right \}
  \\[1ex]
 &\quad+\Tr \, \Tr \left \{
 \frac{\delta^2 \BPsi}{\delta \Psi^T_\Sigma(\bx)\otimes\delta\Psi^T_\Sigma(\bx)}
~\der_\phi \Psi_\Sigma (\bx)
\otimes  \der_\phi \Psi_\Sigma  (\bx) 
\right \}  
+  \frac{\deltab^2 \BPsi }{\deltab \phi(\bx)^2}
 .
\end{split}
\end{align}
\eew
 where $\deltab$ denotes the {\em partial} functional derivative
with respect to $\phi(\bx)$  
and $\delta(\mathbf{0})$ is a regularized value of 
$\delta \Psi_\Sigma (\bx)/ \delta \Psi^T_\Sigma (\bx')$ at $\bx = \bx'$
which can be defined using a point splitting or lattice regularization 
to make sense of $(n-1)$-dimensional delta function $\delta(\bx-\bx')$ at equal points. 
This is the simplest regularization one may use to make sense of the second functional derivative at equal points which appears in the functional derivative Schr\"odinger equation (\ref{fs}).

\subsection{The correspondence between terms $I-V$ in Eq. (\ref{dt}) and  the canonical 
Hamiltonian operator in (\ref{fs})} 
 
\subsubsection{The potential term V}

Our starting observation will be that the term $V$ 
in (\ref{dt}) has to reproduce the 
last 
term in the functional 
derivative Schr\"odinger equation (\ref{fs}). This means that there exists a mapping 
$\mapsto$ such that 
 \bew
\beq \label{map-v}
 \int \!\rd\bx\ 
 \Tr
\left \{
\BPhi(\bx)
~\frac{1}{\ka}
\gamma_0
V(\phi(\bx))\Psi_\Sigma (\bx)) \right \}
\mapsto \int \!\rd\bx \sqrt{-g}\ V(\phi(\bx))~\BPsi , 
\eeq
 \eew
The existence of the map in (\ref{map-v})  implies that 
the following relation should be fulfilled at  any spatial point $\bx$: 
\beq \label{vee}
\Tr \left \{    
\BPhi(\bx) 
 ~ \frac{1}{\ka} \gamma_0 
 \Psi_\Sigma (\bx) \right \}
  \mapsto \sqrt{-g}\ \BPsi .
\eeq
 Then the functional differentiation of both sides of (\ref{vee}) with respect to
$\Psi^T_\Sigma (\bx)$ yields  
 \bew
\beq \label{kade}
\Tr \left \{
\frac{\delta^2 \BPsi}{\delta\Psi^T_\Sigma (\bx)
\otimes \delta\Psi^T_\Sigma (\bx)}
 \frac{1}{\ka} \gamma_0 \Psi_\Sigma (\bx)  \right \}
+ 
 \BPhi(\bx) 
 \frac{1}{\ka}\gamma_0 \delta(\mbox{\bf 0})
 \mapsto
 \sqrt{-g}\ \BPhi(\bx)  , 
\eeq
\eew
where again, 
$\delta(\mathbf{0}) ={\delta \Psi_\Sigma(\bx)}/{\delta \Psi^T_\Sigma(\bx)}.$
 This type of relation is possible 
 if 
 \beq \label{dltpsipsi}
\frac{\delta^2 \BPsi}{\delta\Psi^T_\Sigma (\bx)
 \otimes  \delta\Psi^T_\Sigma (\bx) } =0   
\eeq
 and 
\beq \label{aaa}
\frac{1}{\ka} \gamma_0(x) \delta(\mbox{\bf 0}) -  \sqrt{-g} (x) 
\mapsto 0  
\eeq
The latter {\bf relation} can be understood as the condition 
\beq  \label{lim}
 \gamma^0  \sqrt{-g}   \ka \mapsto \delta(\mbox{\bf 0}) . 
\eeq 
By taking into account that $\sqrt{-g} = \sqrt{- g_{00} h}$, where $h:=\det \Vert g_{ij}\Vert$, 
and $\gamma^0 \sqrt{g_{00}} = 
\underline{\gamma}{}_0$ is the time-like tangent Minkowski space Dirac matrix, 
Eq. (\ref{lim}) can be rewritten as 
\beq \label{limh}
\underline{\gamma}{}_0\ka \mapsto \delta(\mathbf{0})/\sqrt{-h} = \delta^{\mathrm{inv}}(\mathbf{0}) ,
\eeq 
where $\delta^{\mathrm{inv}}(\bx)$ is the invariant $(n-1)$-dimensional delta function  
 defined by the property $\int\! d\bx \sqrt{-h}(\bx)\delta^{inv}(\mathbf{x}) =1$. 
This formula generalizes to curved space-times the limiting map 
$\underline{\gamma}{}_0\ka 
\mapsto \delta(\mathbf{0})$ already found in flat space-time 
\cite{atmp1,atmp2} with the $(n-1)$-dimensional delta function replaced by the invariant one. 
}

\newcommand{\rmvdflat}{ The meaning ... 
 Now, if we recall the origin of Dirac matrices in precanonical quantization as
 the 
 quantum representations of differential forms, we can readily
 recognize the conditions
  (\ref{aaa}) and (\ref{aaaa}) as the inverse quantization map $q$ in Eq.~(\ref{qmap})
 in the limit of
 the 
 infinitesimal elementary volume $\frac{1}{\ka}\rightarrow 0$,
 i.e. Eq. 
 \beq \label{bekade}
\gamma_0\ka\stackrel{q^{-1}}{\longmapsto}\delta(\mbox{\bf 0}) .
\eeq
Note, that one may think of the mapping in Eq. 
 } 

\subsubsection{The second variational derivative term}

 { Our next observation is that 
 the term 
$IV$ in  (\ref{dt}) is able to reproduce the first term 
on the right-hand side of (\ref{dbps2}) in the limiting case ({\ref{lim}}) 
\beq
IV: \quad -\frac{\ka}{2}\gamma_0 \der_{\phi\phi}\Psi_\Sigma 
\mapsto   - \frac{1}{ \sqrt{-g}} g_{00} \delta({\mathbf{0}})\der_{\phi\phi}\Psi_\Sigma .
\eeq  
A comparison with  (\ref{dbps2})  shows that the term $IV$ in  (\ref{dt}) 
 leads to the following expression in functional derivatives of $\BPsi$: 
 \bew
\begin{align} \label{218}
\begin{split} 
IV\!: \; 
\Tr \bigg\{
\frac12 
\BPhi (\bx) 
\varkappa\gamma_0 \der_{\phi\phi}\Psi_\Sigma  (\bx) \bigg\} 
\mapsto 
\frac12 \frac{g_{00}}{\sqrt{-g}}  
\left( \frac{\delta^2 \BPsi }{\delta \phi(\bx)^2} 
 - 2~\Tr \left \{
\frac{\deltab \BPhi(\bx)}{\deltab \phi(\bx)}
~\der_\phi \Psi_\Sigma (\bx) \right \} 
 -  \frac{\deltab^2 \BPsi }{\deltab \phi(\bx)^2} \right)& . 
\end{split}
\end{align}
\eew
The first term on the right-hand side of (\ref{218}) correctly reproduces the  first term in the functional derivative Schr\"odinger equation (\ref{fs}). However, the second and the third term need further investigation.

\subsubsection{The non-ultralocality term and the wave functional $\BPsi$ in terms of precanonical $\Psi$}

Since the right hand side of (\ref{dt}) is expected 
to lead to a functional derivative operator acting on the wave functional $\BPsi$,  
as on the right hand side of the functional Schr\"odinger equation (\ref{fs}), 
the second term in (\ref{218}) with $\der_\phi \Psi_\Sigma$ has to be cancelled by 
the term $II$ in (\ref{dt}) which also contains $\der_\phi \Psi_\Sigma$. 
Therefore, it is required that 
\bew
\beq \label{dypsi} 
II + \mathrm{2nd\,term\,of \,(\ref{218})}:\quad 
 \ri \BPhi^{}({\bx})
 \gamma_0 \gamma^i\der_i\phi(\bx)
 +  \frac{g_{00}}{\sqrt{-g}} 
\frac{\deltab \BPhi (\bx)}{\deltab \phi(\bx)} \mapsto 0, 
\eeq
\eew  
where the sign $\mapsto$ stresses the fact that it is sufficient that the left 
hand side vanishes under the condition (\ref{lim}) rather than as an equality.  
In fact, by  functionally differentiating both sides of (\ref{dypsi})  with respect to  $\phi(\bx')$  
we can see that (\ref{dypsi}) with $\mapsto$ replaced by the equality 
is not an integrable equation in functional derivatives. Nevertheless, 
by bearing in mind  that  (\ref{dypsi}) has to be valid only under the 
condition (\ref{lim}),  the solution for $\BPhi(\bx)$ can be written in the 
form 
\beq \label{bphi}
\BPhi^{}(\bx) =  \BXi([\Psi_\Sigma];\check{\bx})
e^{-\ri \phi(\bx)\gamma^i\der_i\phi(\bx)/\ka},
\eeq
where  the "integration constant" $\BXi([\Psi_\Sigma];\check{\bx})$ is a functional  of $\Psi_\Sigma (\bx')$ 
 on a punctured space with the removed point $\bx$, so that $\bx'\neq \bx$. By construction, this functional  satisfies the identity 
 $$
 \frac{\deltab \BXi([\Psi_\Sigma];\check{\bx})}{\deltab \phi(\bx)}\equiv 0 .
 $$  
Indeed, by differentiating (\ref{bphi}) with respect to $\phi(\bx)$, 
replacing $\ka$ according to the limiting map (\ref{lim}), and taking into account that 
$\gamma_0(\bx)\gamma_0(\bx) =: g_{00}(\bx)$ and $\der_i\delta(\mathbf{0})=0$ (that restricts the admissible class of regularized delta-functions $\delta(\bx$)) we conclude that  (\ref{bphi}) solves (\ref{dypsi}) under the condition (\ref{lim}).  
 Note also that  
  (\ref{bphi}) by construction fulfills 
\beq 
\frac{\delta \BPhi (\bx )}{\delta \Psi^T_\Sigma (\bx )} 
= \frac{\delta^2 \BPsi}{\delta\Psi^T_\Sigma (\bx)
 \otimes  \delta\Psi^T_\Sigma (\bx) } \equiv 0 , 
\eeq
which is consistent with (\ref{dltpsipsi}). 
Thus the required cancellation of the terms with $\der_\phi \Psi_\Sigma(\bx)$ 
(under the condition (\ref{lim})) fixes the form of the functional $\BPhi(\bx)$ 
introduced in (\ref{ffi}). 
 %
 This allows us to express the wave functional $\BPsi$  
 in the form 
\bew
\beq \label{bpsi3}
\BPsi \sim  
 \Tr
\left \{\BXi([\Psi_\Sigma];\check{\bx})~
e^{-\ri \phi(\bx)\gamma^i\der_i\phi(\bx)/\ka}~
 \frac{\gamma_0}{\sqrt{-g} \ka}
 \Psi_\Sigma (\bx) \right \}_{\mbox{\Large $\rvert$} \scriptstyle
 \ka \stackrel{\mbox{\tiny $$}}{\mbox{\scriptsize $\longmapsto$}} 
 \gamma_0\delta(\mathbf{0})/ \sqrt{-g}  },
\eeq  
\eew
valid at any point  $\bx$. The equality up to a normalization factor  which will depend on  $\ka$ and $\sqrt{-g}$ 
is denoted as  $\sim$ 
The notation $\{ ... \}_{\mbox{\large $\rvert$} \scriptstyle
 \ka \stackrel{\mbox{\tiny $$}}{\mbox{\scriptsize $\longmapsto$}} 
 \gamma_0\delta(\mathbf{0}) / \sqrt{-g}}$
 indicates that every appearance of $\ka$ in the expression inside braces 
 is replaced  by  
 $\gamma_0\delta(\mathbf{0}) / \sqrt{-g}$ 
 as prescribed by the limiting map (\ref{lim}). 

Using 
(\ref{bpsi3})
we can now evaluate the last term 
in (\ref{218}) in the limit (\ref{lim}): 
 \bew
\beq 
\mbox{3-rd term of (\ref{218}):}
\quad 
\frac12 \frac{g_{00}}{\sqrt{-g}} \frac{\deltab^2 \BPsi }{\deltab \phi(\bx)^2}
\mapsto
 -  \frac{1}{2}\sqrt{-g} g^{ij}\der_i \phi(\bx) \der_j \phi(\bx) \BPsi .  
\eeq
\eew
The right hand side  correctly reproduces the second term in the functional derivative Schr\"o\-dinger equation (\ref{fs}), thus correctly accounting for the inherent non-ultralocality 
of  quantum relativistic scalar field theory (cf. \cite{kld}) in curved space-time. 

}

{ Thus, 
all terms in the functional derivative Schr\"odinger 
equation (\ref{fs}) are now derived from 
pSE restricted to $\Sigma$, eq. (\ref{eq6n}).  
However, there are still unaccounted for  terms $I$, $IIIa$ and $IIIb$  in (\ref{dt}) 
 \bew
 \beq \label{227}
 I+IIIa+IIIb:\quad -\ri \int \!\rd\bx\ \Tr \Big\{\BPhi(\bx) \gamma_0\gamma^i \nabla^{tot}_i\Psi_\Sigma 
 + \BPhi(\bx) [\omega_0,\Psi_\Sigma]
  \Big\} .
 \eeq
 \eew
In flat space-time \cite{atmp1,atmp2,romp2018}, it is the corresponding term 
 with 
 $d \Psi_\Sigma (\bx)/ d x^i $ 
vanishes if 
$\Psi_\Sigma (\bx)$ vanishes at the spatial infinity. 
Let us see if or how this property extends to  curved space-times.

\subsubsection{The vanishing contribution from the terms I and IIIa }

At first we consider the first term in (\ref{227}). 
 Using the covariant Stokes theorem 
  from (\ref{227}) we obtain 
\bew
\begin{align} \label{first} 
\begin{split}
I+IIIa:\quad 
&-\ri \int \!\rd\bx\ \Tr \Big\{\BPhi(\bx) \gamma_0\gamma^i \nabla^{tot}_i\Psi_\Sigma \Big\} 
=  
-\ri \int \!\rd\bx\ \sqrt{-h}\
\left( \Tr \Big\{ \frac{1}{\sqrt{-h}}
 \BPhi(\bx) \gamma_0\gamma^i \nabla^{tot}_i\Psi_\Sigma \Big\}  
 \right) 
\\ = & 
-\ri \int \!\rd\bx\ \sqrt{-h}  \Tr 
\Big\{ \nabla^{tot}_i \Big(\frac{1}{\sqrt{-h}}
 \BPhi(\bx) \gamma_0\gamma^i  \Psi_\Sigma \Big)\Big\}  
  + \ri \int \!\rd\bx\ \left( \sqrt{-h} \Tr 
\Big\{\nabla^{tot}_i \Big(\frac{1}{\sqrt{-h}}
 \BPhi(\bx) \gamma_0\gamma^i\Big) \Psi_\Sigma \Big\} 
 \right)
 \\
 = & -\ri \oint_{\der \Sigma} \!\rd\bx_i 
  \Tr \Big\{ \BPhi \gamma_0\gamma^i \Psi_\Sigma \Big\} 
+  \ri \int \!\rd\bx\  \Tr 
\Big\{ \BPhi \big(\nabla^{tot}_i ( \gamma_0\gamma^i) \big) \Psi_\Sigma \Big\}
 + \ri \int \!\rd\bx\ 
 \left( \frac{-\nabla_i \sqrt{-h}}{\sqrt{-h}}
\Tr \Big\{ \BPhi  \gamma_0\gamma^i\Psi_\Sigma \Big\} 
\right. \\ & +  \Tr 
\Big\{ \Big(\nabla^{tot}_i 
 \BPhi(\bx) \Big) \gamma_0\gamma^i \Psi_\Sigma \Big\} 
 \bigg) . 
  \end{split}
  \end{align} 
\eew 
where $\rd \bx_i = d^{n-2} \bx|_{\der\Sigma} n_i (\bx) $ 
is the measure of $(n-2)$-dimensional integration over the boundary $\der\Sigma$ with the normal vector 
$n_i (\bx)$ tangent to $\Sigma$.  On the right hand side of (\ref{first}), 
\begin{itemize}
\item[(i)]
the first boundary term 
is the result of the covariant Stokes theorem and it vanishes if 
$\Psi$ vanishes on the boundary $\der \Sigma$;

\item[(ii)]  
 The following three terms 
 follow from the Leibniz rule for the total 
covariant derivative $\nabla^{tot}_i$ with respect to the Clifford products of 
tensor Clifford-algebra-valued functions. 

\item[(iii)] 
In the second term, $\nabla^{tot}_i ( \gamma_0\gamma^i) = 0$
 due to the covariant constancy of Dirac matrices (\ref{mcomp}). 

\item[(iv)]
In the third term, 
the metric compatibility yields $\nabla_i \sqrt{-h} = 0$.  

\item[(v)] 
In the fourth term,  
the explicit formula for $\BPhi (\bx) $ in (\ref{bphi}) 
 yields 
\bew
\beq \label{229}
\nabla^{tot}_i \BPhi (\bx) = \frac{-\ri}{\ka}\BPhi (\bx) \big(\der_i \phi\gamma^l\der_l \phi 
+ \phi \gamma^l \der_{il} \phi + \phi (\nabla^{tot}_i \gamma^l) \der_l \phi \big).
\eeq
\eew
 Noticing that the last term in (\ref{229}) vanishes due to (\ref{mcomp})  
 and substituting  (\ref{229}) into the last term in (\ref{first}), 
 using the covariant Stokes theorem  
 and the assumption that the field configurations $\phi(\bx)$ vanish  at the spatial infinity, we obtain 
\bew
\begin{align} 
\begin{split} \label{230}
 \int \!\rd\bx\ \Tr \Big\{   
 \BPhi(\bx)  \frac{1}{\ka} \gamma_0 \Psi_\Sigma (\bx) 
\big( g^{il}\der_i \phi \der_l\phi + \phi g^{il}\der_{il}\phi \big)  \Big\} 
& = \int \!\rd\bx\ \Tr \Big\{   
 \Vert\BPsi\Vert \sqrt{-g} 
\big( g^{il}\der_i \phi \der_l\phi + \phi g^{il}\der_{il}\phi \big)  \Big\} 
\\
& = 
- \BPsi \int \!\rd\bx\ \sqrt{-h} \nabla^{}_i\big(\sqrt{g_{00}}g^{il}\big)\frac12 \der_l \phi^2
=0,  
\end{split}
\end{align}
\eew
where we use the fact that the matrix-valued functional 
$\Vert\BPsi\Vert :=  \BPhi(\bx)  \frac{1}{\ka \sqrt{-g}} \gamma_0 \Psi_\Sigma (\bx)$ 
such that  $\Tr  \Vert\BPsi\Vert = \BPsi$ (c.f. (\ref{vee})) is independent of $\bx$.    
The result is that the right-hand side of (\ref{230}) vanishes because of the metricity 
of space-time: 
$\nabla^{}_\alpha g^{\mu\nu} = 0$. 
\end{itemize}

\medskip 

Therefore, it is demonstrated that in the limiting case (\ref{lim}) all four terms on the right-hand side of (\ref{first}) 
vanish so that the terms $I$ and $IIIa$ in (\ref{dt}) do not  contribute to the equation for the functional $\BPsi$.  
 \newcommand{\rmvdfromolderstatic}{ 
Note that this result is a consequence of the properties of the pseudo-Riemannian geometry of the space-time background,
the assumed boundary conditions that the values of $\Psi_\Sigma(\bx)$ and $\phi(\bx)$ at $\bx \rightarrow \infty$ are vanishing, and the particular form of the functional $\BPhi(\bx)$  which was established earlier in (\ref{bphi}).  The covariant Stokes theorem and the 
Leibniz property of the covariant derivative with respect to the Clifford product 
have been instrumental. 
The latter property is guaranteed only when the 
spin-connection matrix acts on the Clifford-algebra-valued wave functions by the 
commutator product. 
 } 
By combining the above considerations 
we  obtain from (\ref{dt}) the following equation for the functional $\BPsi$: 
\bew
\beq \label{fs1}
\ri\hbar \der_t \BPsi = 
 \int \! d\bx\, \sqrt{-g}
 \left (  \frac{\hbar^2}{2}\ \frac{g_{00}}{g}\frac{\delta^2}{\delta \phi(\bx)^2} 
- \frac{1}{2} g{}^{ij} \der_i\phi(\bx)\der_j\phi(\bx) + V(\phi)
 \right) \BPsi 
 - \frac\ri4 \Tr \Big\{ \BPhi (\bx)  [\omega_{0\Sigma}, \Psi_\Sigma] \Big\} . 
\eeq 
\eew 
We see that the first three terms in the right hand side reproduce the canonical Hamiltonian operator in the 
functional derivative Schr\"odinger equation (\ref{fs}).  However, the last term, which does not vanish 
in arbitrary non-static space-times, 
still can not be expressed in terms of $\BPsi$. For this reason, we will treat static and non-static space-times separately.

\subsection{Static space-times}

 In static space-times when $\omega_0=0$  equation (\ref{fs1}) 
coincides with the canonical functional derivative Schr\"odinger equation (\ref{fs}). 
Thus the latter is derived from the precanonical Schr\"odinger equation as the limiting 
case corresponding to (\ref{lim}). In this case,  we can also specify the functional 
$\BXi([\Psi_\Sigma(\bx)], \check{\bx})$ in (\ref{bpsi3}) by combining the observations 
presented above together and noticing that the relation (\ref{bpsi3}) is valid at 
{\it any\/} given point ${\bx}$.
This 
is possible only if the functional $\BPsi$ 
 is the continuous product of identical terms at all points ${\bx}$, namely  
\bew
\beq\label{schrod}
\BPsi \sim 
  \Tr \bigg \{\prod_\bx
e^{-i\phi(\bx)\gamma^i\der_i\phi(\bx)/\ka} 
 \underline{\gamma}{}_0
 \Psi_\Sigma (\phi(\bx), \bx, t)
\bigg\}{}_{\mbox{\Large $\rvert$} \scriptstyle
 \ka 
 \mapsto 
 \gamma_0\delta(\mathbf{0})/ \sqrt{-g}  } , 
\eeq 
\eew
where $\sim$ means an equality up to a normalization 
factor which includes $\ka$ and $\sqrt{-h}$. 
The formal continuous product expression in (\ref{schrod}) 
 can be understood as a multidimensional product integral \cite{prodint} 
(see (\ref{limh})) 
\bew
\beq \label{pint}
\BPsi \sim 
  \Tr \left \{   
  \underset{\!\bx}{\scalebox{1.5}{$\displaystyle \prodi$}} 
  e^{-i\phi(\bx)\gamma^i (\bx) \der_i\phi(\bx)/\ka}
\underline{\gamma}{}_0 \Psi_\Sigma (\phi(\bx), \bx, t)
\right \}{}_{\mbox{\Large $\rvert$} \scriptstyle
 \frac1\ka \underline{\gamma}{}_0
 \mapsto 
 \sqrt{-h}\rd\bx  }, 
\eeq  
\eew
This expression generalizes a similar  result obtained in flat space-time \cite{atmp2}.  
The only difference is that in curved space-time  the spatial integration 
measure $\rd \bx$ is replaced by the invariant one $\sqrt{-h}\rd\bx$ 
and the Dirac matrices are $\bx$-dependent. In $(1+1)$-dimensional space-time 
the product integral above is given by the well knows path-ordered exponential or the Peano-Baker series. 
A multidimensional generalization is briefly discussed in the books cited in \cite{prodint} and probably needs further refinement. 
However, in our case, only the trace of the product integral of Clifford-algebra valued functions rather than arbitrary non-commutative 
matrices has to be defined, which simplifies the task of defining the expression (\ref{pint}): for example, under the trace each 
term in the Peano-Baker series defining the one-dimensional product integral is a cyclic permutable product of matrices 
rather than arbitrarily noncommutative.

By taking into account the fact that some of the terms in (\ref{first}) are proven to not contribute to the time evolution of $\BPsi$  
we can write the effective equation which govern the time evolution of $\Psi_\Sigma $  which does contribute to the time 
evolution of the wave functional $\BPsi$: 
\beqa \label{e37}
\ri\der_t \Psi_\Sigma &=  \gamma_0 \left( -\frac\ka2\der_{\phi\phi} 
+ \ \ri\gamma^i \der_{i} \phi (\bx)\der_\phi 
+ \frac1\ka V(\phi)) \right)  \Psi_\Sigma 
\nn \\ & 
=: \gamma_0 \mathcal{E}
\eeqa
By substituting $\Psi_\Sigma $ in the form 
\beq\label{e40}
\Psi_\Sigma = 
e^{+\frac{\ri}{\ka} \phi(\bx)\gamma^i \der_i\phi(\bx)} 
\Phi_\Sigma , 
\eeq  
we obtain 
\beq
\ri\der_t \Psi_\Sigma = e^{+\frac{\ri}{\ka} \phi(\bx)\gamma^i \der_i\phi(\bx)} 
\ri\der_t \Phi_\Sigma 
\eeq
on the left hand side of (\ref{e37}) and 
\bew
\beq
\gamma_0e^{+\frac{\ri}{\ka} \phi(\bx)\gamma^i \der_i\phi(\bx)} 
\left( -\frac\ka2\der_{\phi\phi} - \frac{1}{2\ka}g^{ij} (\bx) \der_i \phi (\bx) \der_j \phi (\bx) + \frac1\ka V(\phi) \right)  \Phi_\Sigma 
\eeq
\eew
on the right hand side. Hence, 
\bew
\beq
\ri\der_t \Phi_\Sigma = \gammat_0 (\bx) 
\left( -\frac\ka2\der_{\phi\phi} - \frac{1}{2\ka}g^{ij} (\bx) \der_i \phi (\bx) \der_j \phi (\bx) + \frac1\ka V(\phi) \right)  \Phi_\Sigma 
\eeq
\eew 
where 
\beq \label{eq42}
\gammat_0 (\bx):=
e^{-\frac{\ri}{\ka} \phi(\bx)\gamma^i \der_i\phi(\bx)} 
\gamma_0 (\bx) e^{+\frac{\ri}{\ka} \phi(\bx)\gamma^i \der_i\phi(\bx)} 
\eeq 
Obviously, $\gammat_0 (\bx) \gammat_0 (\bx) = \gamma_0 (\bx)\gamma_0 (\bx)=g_{00}(\bx)$, 
hence the transformation in (\ref{eq42}) is a Clifford algebra isomorphism. 


\subsection{Non-static space-times}

In non-static space-times, when $\omega_0\neq 0$, the last term in (\ref{fs1}) does not allow us to obtain a close equation for the functional $\BPsi$. In order to find a way out,  
let us write the effective equation similar to (\ref{e37}) which governs the time evolution of 
$\Psi_\Sigma$, with the term $I$ and the spatial part of the term $IIIa$ in (\ref{dt}),
  which are proven in (\ref{first}) to have no contribution to $\der_t\BPsi$,    removed: 
\bew
\begin{align}\label{231}
\ri\der_t \Psi_\Sigma = \gamma_0  \left( -\frac\ka2\der_{\phi\phi} 
+ \ri\gamma^i \der_{i} \phi (\bx) \der_\phi 
+ \frac1\ka V(\phi)) \right) \Psi_\Sigma  
- \frac{\ri}{4}  [\omega_0{}_\Sigma, \Psi_\Sigma ]
= : \hat{H}{}_0 -  \frac{\ri}{4}  [\omega_0{}_\Sigma, \Psi_\Sigma ] . 
\end{align} 
\eew
We first note that by transforming $\Psi_\Sigma$ as follows: 
\beq
\Psi_\Sigma := U \Psi'_\Sigma U^{-1} , 
\eeq
where 
\beq \label{e36}
U(\bx,t):= [e^{ - \frac14 \int_0^t ds\ \omega_0(\bx,s) }]
\eeq
is the tranformation determined by the path-ordered exponential inside the square brackets, 
we obtain 
\beq
\ri \der_t \Psi = - \frac\ri4[\omega_0, \Psi] + U \ri \der_t \Psi'_\Sigma U^{-1} .
\eeq
Then 
\beq \label{52}
\ri\der_t \Psi'_\Sigma = U^{-1} H_0 \Psi_\Sigma U 
= H_0' \Psi' , 
\eeq 
where $ \Psi' := U^{-1}\Psi_\Sigma U$, $H_0':= U^{-1} H_0 U$. 
As the transformation $U$ affects only the terms with $\gamma^\mu-$s, 
\beq
\hat{H}_0{}' 
= \gamma'_0  \left( -\frac\ka2\der_{\phi\phi} 
+ \ri\gamma'{}^i \der_{i} \phi (\bx) \der_\phi 
+ \frac1\ka V(\phi)) \right) , 
\eeq
where 
\beq
\gamma'{}^\mu (\bx,t) := U^{-1}(\bx,t) \gamma^i (x) U (\bx,t) .
\eeq
It is easy to check that  
\beq
\gamma'{}^\mu\gamma'{}^\nu + \gamma'{}^\nu \gamma'{}^\mu = 2 U^{-1}g^{\mu\nu}U 
= 2 g^{\mu\nu} . 
\eeq
Hence  the $U$-transformation is just a rotation in the Clifford algebra of space-time. 

Using (\ref{52}) one can write 
\begin{align}
\ri \der_t \BPsi &= \Tr \int\! d\bx\ \frac{\delta \BPsi}{\delta \Psi'{}^T_\Sigma (\bx)}  \ri \der_t \Psi'_\Sigma 
\\ & = \Tr \int\! d\bx\, \frac{\delta \BPsi}{\delta \Psi'{}^T_\Sigma(\bx)} \hat{H}'_0 \Psi'_\Sigma .
\end{align}
By comparing it with (\ref{dtps})  and (\ref{eq6n}) we conclude that the results 
obtained in static space-times with  can be generalized to 
curved space-times with $\omega_0 \neq 0$ using the dictionary: 
\beqa
\gamma^\mu &\rightarrow\ \gamma'^\mu &=\ U^{-1}\gamma^\mu U ,\\ 
\Psi_\Sigma &\rightarrow\ \Psi'_\Sigma &=\ U^{-1} \Psi_\Sigma U , \\
\hat{H}_{0\Sigma} &\ \rightarrow\ \hat{H}{}'_{0\Sigma} &=\ U^{-1}H_{0\Sigma} U 
\eeqa
with $U$ given by (\ref{e36}). 
Then, the wave functional (\ref{pint}) rewritten in terms of the primed objects:   
\bew
\beq \label{e60}
\BPsi \sim 
 \Tr \left \{   
  \underset{\!\!\bx}{\scalebox{1.5}{$\displaystyle \prodi$}} 
  e^{-i\phi(\bx)\gamma'{}^i (\bx,t) \der_i\phi(\bx)/\ka}
\underline{\gamma}{}_0 \Psi'_\Sigma (\phi(\bx), \bx, t)
\right \}{}_{\mbox{\Large $\rvert$} \scriptstyle
 \underline{\gamma}{}_0\frac1\ka 
 \mapsto 
 \sqrt{-h}\rd\bx  }  
\eeq
\eew  
represents, up to a normalization factor, the Schr\"odinger wave functional in terms of precanonical wave functions 
in an arbitrary curved space-time (whose metric is represented in Gaussian coordinates 
with $g_{0i} = 0$) and it 
satisfies (\ref{fs1}) without the last term, i.e. the functional derivative 
Schr\"odinger equation (\ref{fs}).


\medskip

In such a way we have demonstrated that in curved space-time the precanonical Schr\"odinger equation (\ref{crv-ns}) 
allows us to obtain the canonical 
functional derivative Schr\"odinger equation (\ref{fs})  and the explicit product integral formula (\ref{pint}) 
relating the Schr\"odinger wave functional with the Clifford-valued 
precanonical wave function 
in the 
limiting case when 
$\underline{\gamma}{}_0\ka$  is replaced by 
$\delta{\bf(0)}/\sqrt{-h}$, (a regularized) invariant delta-function 
which can be interpreted as the UV cutoff of the total volume of the momentum space 
which one has to introduce in order to make sense of the second variational derivative at equal points 
in (\ref{fs}). As  in the previously considered  case of flat space-time, the standard 
formulation of QFT in Schr\"odinger functional representation emerges from the precanonical description 
as a singular limiting case.

}

\section{Conclusion}

We explored a  connection between the description of an interacting quantum 
scalar field in curved space-time derived from precanonical quantization 
and  the standard description in the functional picture resulting from the canonical quantization. 

We have demonstrated that the functional derivative Schr\"odinger equation (\ref{fs}) can be derived from the 
partial derivative precanonical Schr\"odinger equation (\ref{crv-ns})
in the limiting case (\ref{lim}). Namely, the restriction of the precanonical Schr\"odinger equation to 
the subspace $\Sigma$ representing a field configuration at time $t$, eq. (\ref{eq6n}), governs the time evolution of the 
wave functional according to (\ref{dtps}) and (\ref{dt}). Then, in the limiting case (\ref{lim}), 
\begin{itemize}

\item[(i)] the potential term $V$ in (\ref{dt}) reproduces the potential term in (\ref{fs}); 

\item[(ii)] the term $IV$ in (\ref{dt}) reproduces the second functional derivative term 
 in (\ref{fs}) up to some  additional terms which have no obvious counterpart in (\ref{fs});
 
\item[(iii)] by noticing that 
    one of those additional terms  can be cancelled by the term $II$ in (\ref{dt})
 we obtain an 
expression of the Schr\"odinger wave functional as a trace of the continuous product  
of the precanonical wave functions (\ref{schrod}) which we suggested to interpret as a multidimensional 
analogue of the product integral, eq. (\ref{pint}); 
 
\item[(iv)] using the  expression of the wave functional in terms of pre3canonical wave functions 
  in the other additional term mentioned in (ii) we 
  reproduce the second term on the right hand side of (\ref{fs});  

\item[(v)] this explicit expression also allows us to show that 
for  the fields $\phi(\bx)$ and $\Psi_\Sigma(\phi(\bx),\bx,t)$ vanishing at the spatial infinity
the remaining terms $I$ and  $IIIa$ do not contribute to the functional Schr\"odinger equation 
(\ref{fs}); 

\item[(vi)] in static space-times when $\omega_0 = 0$ the remaining term $IIIb$ vanishes and 
the functional Schr\"odinger equation (\ref{fs}) emerges from the precanonical Schr\"odinger equation (\ref{crv-ns}) 
and the Schr\"odinger wave functioneal is expressed in terms of precanonical wave functions; 

\item[(vii)] In non-static space-times with $\omega_0 \neq 0$,  we argue that 
the transformation  (\ref{e36}) absorbs the contribution of the  term  $IIIb$ 
in (\ref{dt}) thus allowing us to 
obtain the functional Schr\"odinger equation 
(\ref{fs}) from the precanonical Schr\"odinger equation (\ref{crv-ns}) 
and to express the Schr\"odinger wave functional in terms of transformed precanonical wave functions, eq. (\ref{e60}).  
\end{itemize}

These results generalize to arbitrary curved  space-times  
the statement from 
\cite{atmp1,atmp2,romp2018}
that the standard functional Schr\"odinger representation of 
QFT is a certain  (symbolic) limiting case of the theory of quantum fields obtained by precanonical quantization. 

The symbolic or singular nature of the limiting transition to the standard formulation of QFT in functional Schr\"odinger representation 
is related to the fact that the latter, due to the presence of the second functional derivative in coinciding points, is not a
well-defined theory unless a regularization is introduced. The latter introduces a UV cutoff scale as an additional element of the theory 
removed by a subsequent renormalization. In precanonical quantization, a UV scale $\varkappa$ appears as an inherent element quantization. 
However, in so doing, it does not alter the relativistic space-time at smaller scales. 

Whether $\varkappa$ is a fundamental scale or an auxiliary element of precanonical quantization of fields is still an open question. 
On the one hand,  one can show that in free scalar theory $\varkappa$ disappears from the observable characteristics of a qauntum field 
and in interacting scalar theory $\varkappa$ enters in the perturbative corrections to the mass spectrum, thus suggesting it 
can be renormalized away by replacing the expressions in terms of a bare mass and $\varkappa$ by the "observed mass". On the 
other hand, an estimation of the mass gap in the  quantum pure SU(2) gauge theory \cite{ym-mass} derived by 
precanonical quantization  and a naive estimation of the cosmological constant based on the precanonically quantized pure Einstein gravity \cite{qg} seem to consistently point to the estimation of the scale of $\varkappa$ at $\sim 10^2 MeV$.  We hope to clarify this issue 
in our forthcoming work. 
 
 \bigskip

 \paragraph*{Acknowlegdements:} 
I gratefully appreciate the hospitality of the School of Physics and Astronomy 
of the University of St Andrews, Scotland, and 24/7 availability of its facilities for 
research.


{ 

 }

\end{document}